\documentclass[11pt]{article}

\usepackage[margin=1in]{geometry}
\usepackage{amsmath,amssymb,amsthm}
\usepackage{graphicx}
\usepackage{enumitem}
\usepackage{hyperref}
\usepackage{float} 
\title{Indefinite Causal Order from Failure-to-Glue: \\Contextual Semantics and Parametric Time}

\author{Partha Ghose\footnote{partha.ghose@gmail.com} \\
Tagore Centre for Natural Sciences and Philosophy,\\ Rabindra Tirtha, New Town, Kolkata 700156, India}
\date{}

\newtheorem{definition}{Definition}

\newtheorem{proposition}{Proposition}
\newtheorem{example}{Example}

\begin{document}
\maketitle
\begin{abstract}
Indefinite causal order (ICO) has been approached from several operational and conceptual directions,
including higher-order quantum processes (such as the quantum switch), the process-matrix framework,
and quantum-gravity motivated proposals involving superposed causal structure. Yet the meaning of
``indefiniteness'' and its relation to definite-order explanations often remain conceptually opaque.

This paper develops a unified semantic--structural framework in two parts. In Part~I, we adopt a
category-theoretic viewpoint in which each definite causal ordering (a partial order or directed acyclic
graph (DAG) type) is treated as a \emph{context}. We then ask a simple question: can the observed behaviour
be explained by consistently stitching together definite-order descriptions (perhaps by mixing different
definite orders from run to run)? Formally, this ``stitching'' requirement is expressed as a \emph{gluing}
condition: causal separability corresponds to the existence of a global section (a consistent global choice)
for a suitable family of admissible definite-order descriptions, while causal nonseparability is a
failure-to-glue. To articulate causal statements across incompatible contexts, we introduce a compact
seven-valued contextual classifier as an explicit elaboration of intuitionistic semantics, distinguishing
mere variation across contexts from genuine indeterminacy.

Part~II applies this framework to a quantum-gravity motivated setting in which time is taken to be a
fundamental \emph{parametric} ordering variable $\tau$, distinct from geometric (spacetime) time. Adopting a
stochastic-quantization perspective on spin-network dynamics---in which Hilbert space is not assumed
fundamental and the Wheeler--DeWitt condition is read as an equilibrium/stationarity constraint---we argue
that ICO is most naturally understood as indefiniteness in the \emph{parametric ordering} of coarse-grained
relational interventions, even when the underlying microscopic update process is globally ordered by $\tau$.
Taken together, the two parts provide a precise language for this form of order-indefiniteness and clarify
how operational ICO, higher-order processes, and emergent spacetime structure fit together without conflation.
\end{abstract}
\bigskip
\begin{center}
\large\bfseries Part I\\
\medskip
Failure-to-Glue and Seven-Valued Contextual Semantics
\end{center}
\bigskip
\section{Introduction}

In most physical reasoning---and in essentially all textbook models of computation---the order of events is taken to be definite: causes precede effects, algorithms are sequences, and experimental procedures are circuits with a fixed `before/after' structure. Yet quantum theory invites us to revisit what is meant by `order' once relevant degrees of freedom can be placed in coherent superposition, and quantum-gravity considerations suggest that causal structure might not be a fixed background property at all. A particularly lucid conceptual introduction to such indefinite causal order (ICO) senarios is the recent `map' of the arena   by Escand\'on-Monardes \cite{EscandonMonardes2025}, which organises the field into a small number of interlocking threads (from Hardy's programme to quantum switches, process matrices, and superpositions of causal structure) and culminates in a clear list of ongoing debates and open problems.

\subsection*{Hardy's programme: probabilistic physics with non-fixed causal structure}

A central motivation for ICO comes from the tension between general relativity and quantum theory. General relativity is deterministic but its causal structure is dynamical (it depends on the distribution of matter and energy), whereas ordinary quantum theory is fundamentally probabilistic but typically presupposes a fixed causal background for laboratory operations. Hardy argued that an eventual quantum-gravity framework should therefore combine these `radical' features, yielding a probabilistic theory with non-fixed causal structure \cite{Hardy2007}. In a subsequent work he discussed the consequences of indefinite causal structure for computation and asked whether such indefiniteness can function as a computational resource \cite{Hardy2009}.

In much of the quantum-gravity-motivated ICO literature, including Hardy's line of thought and later ``superposed causal structure'' scenarios, the indefiniteness under discussion is tied to \emph{geometric} causal relations: one considers alternative spacetime causal structures (or geometries) and, at least conceptually, their coherent superposition. Such discussions are typically framed in (or translated into) a Hilbert-space language in which superpositions of causal relations or geometries are meaningful. In that setting, ``time'' is geometric (spacetime) time---even if its global ordering properties are non-classical. The present paper keeps this strand in view, but Part~II adopts an \emph{alternative} stance:
working within a stochastic-mechanical foundation in which Hilbert space is not taken as fundamental,
we propose that the time relevant to ICO in quantum gravity is a \emph{parametric} ordering variable $\tau$,
with geometric time and spacetime causal cones arising only in an emergent semiclassical regime
\cite{NandiGhosePetruccione2025}.

\subsection*{Three operational strands: switches, process matrices, and superposed causal structures}

Following \cite{EscandonMonardes2025}, it is useful to distinguish three operational strands.

\paragraph{(i) Quantum switch.} The quantum switch is a paradigmatic higher-order operation: it takes two (or more) quantum operations as inputs and outputs an effective transformation in which their order is coherently controlled by an ancillary system. This viewpoint is naturally expressed in the language of supermaps/combs and higher-order maps \cite{Chiribella}. The switch has motivated a large body of work on quantum-information advantages and has become the most accessible concrete exemplar of order-indefinite behaviour.

\paragraph{(ii) Process matrices.} The process-matrix framework makes a sharper separation: inside each local laboratory, parties implement ordinary quantum operations, but no global causal order is assumed between laboratories. The connecting object is a process matrix, which determines the joint probabilities generated by the parties' local actions \cite{OCB}. A key structural notion is causal (non)separability: a process is causally separable if it can be expressed as a convex mixture of definite-order processes; otherwise it is causally nonseparable. In some cases incompatibility with definite causal order can be witnessed device-independently via violations of causal inequalities \cite{OCB}.

\paragraph{(iii) Superposition of causal structures.} A third strand takes the quantum-gravity motivation more literally by considering superpositions of spacetime geometries (or, more generally, superpositions of causal structures). In such scenarios the causal relation between two events may depend on the branch of the superposition, and one obtains a superposition of causal orders. Concrete `gravitational switch' scenarios linking order-indefiniteness to superposed gravitational fields have been proposed \cite{Zych2019}.

\begin{figure}[H]
  \centering
  \includegraphics[width=\textwidth]{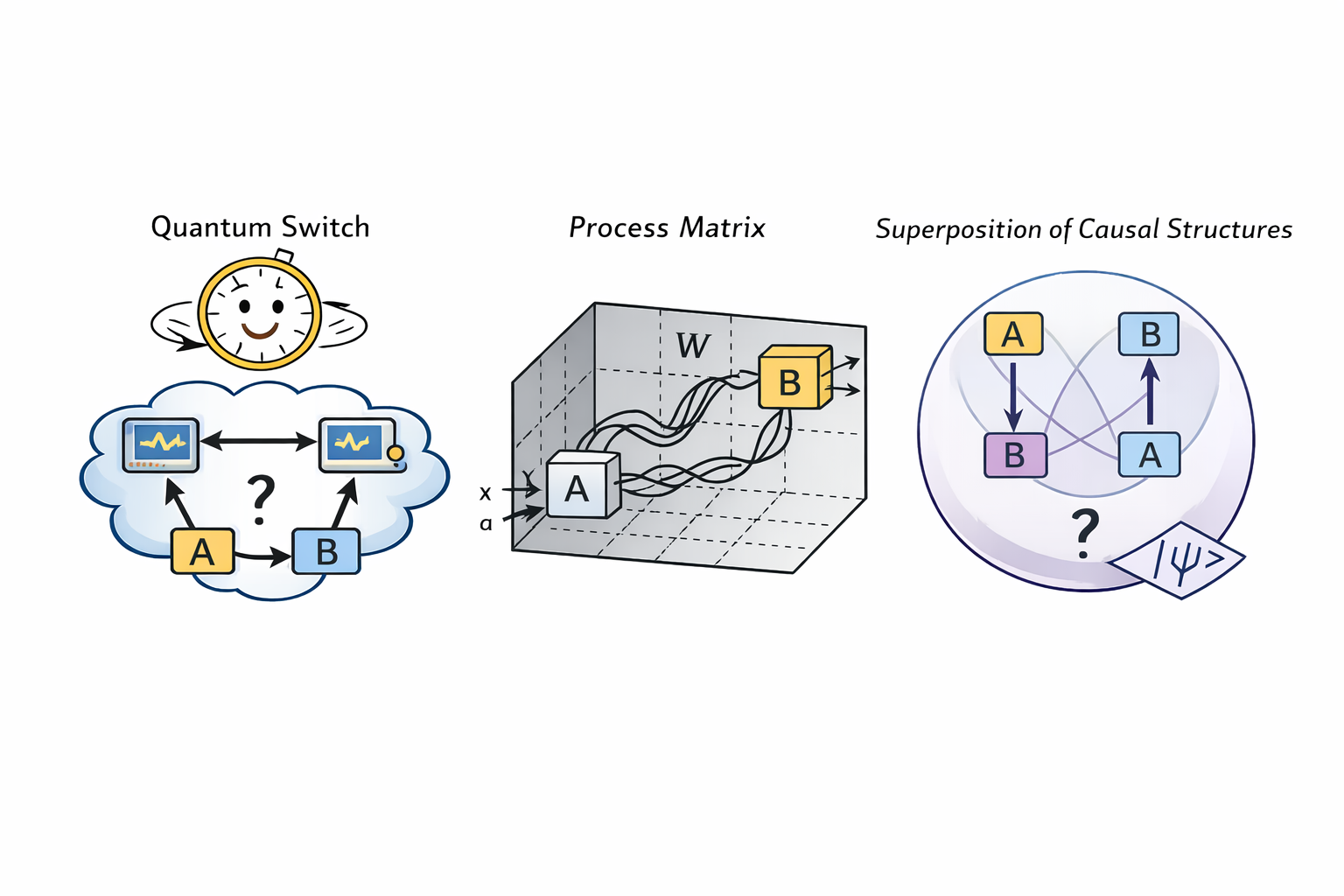}
  \caption{Cartoon roadmap: (left) quantum switch (coherent control of order), (middle) process-matrix framework (a global process $W$ connecting local labs), and (right) superposition of causal structures (branch-dependent causal relations).}
  \label{fig:ico-cartoons}
\end{figure}

\subsection*{Debates and open problems}

Even at a conceptual level, the ICO landscape comes with persistent open questions: what is the operative resource behind observed advantages (order-indefiniteness itself, coherence of a control system, non-commutativity, or combinations thereof); where precisely the boundary lies between physically implementable and merely formal process-matrix objects; and how to make precise the notion of events and localisation in `superposed spacetime' scenarios. A clear overview of these debates and a list of open problems is given in \cite{EscandonMonardes2025}.

\subsection*{What this paper adds: failure-to-glue and seven-valued contextual semantics}

This paper does not add a \emph{fourth strand} to the ICO literature. Rather, it proposes a
semantic--structural unification designed to clarify the existing landscape, especially for readers
not already committed to a single formalism. Mathematically, we use a category-theoretic language (categories, presheaves, and global sections) to formulate ‘definite-order explainability’ as a gluing problem.

\begin{enumerate}[leftmargin=2em]
\item \textbf{Failure-to-glue (no global section).} We treat definite causal structures (partial orders/DAG-types of laboratories) as contexts organised into a category (or poset) of causal-order contexts. Over this base we consider a convex presheaf of admissible definite-order models. In this setting, causal separability becomes a globalisability statement: the existence of a global section implementing a definite-order explanation (possibly as a classical mixture). Conversely, causal nonseparability becomes a canonical failure-to-glue of definite-order sections into a global one. This places ICO alongside the sheaf-theoretic 'no global section' paradigm familiar from contextuality \cite{AB,ButterfieldIsham1998,Isham2010ToposMethods}.

\item \textbf{Seven-valued semantics as an intuitionistic elaboration.} Within each fixed context we use ordinary intuitionistic forcing (warranted assertion) for causal propositions. Across a family of contexts, we summarise the status of a proposition by whether it is supported somewhere, refuted somewhere, and/or indeterminate somewhere. The seven non-empty combinations yield a compact seven-valued contextual classifier. This does not replace intuitionistic logic; it elaborates it into a disciplined metalanguage for talking across incompatible causal contexts.
\end{enumerate}

Finally, the framework interfaces naturally with higher-order causal categories: such categories provide an internal compositional logic of admissible wiring and higher-order maps \cite{Simmons,Wilson,KU,Bisio}, while the present paper supplies a contextual semantics for statements about causal order and its (in)definiteness.

\section{Contexts: definite causal structures as a base}

\begin{definition}[DAG and definite causal contexts]
A \emph{DAG} (directed acyclic graph) is a directed graph with no directed cycle.
In causal modelling, a DAG encodes a definite causal structure: an arrow $X\to Y$ indicates that
$X$ may influence $Y$. Acyclicity rules out time-travel-like causal loops.
\end{definition}

\begin{definition}[Order-context category/poset $\mathcal{O}$]
Let $\mathcal{O}$ be a category (often a poset) whose objects are \emph{definite} causal structures
(order-types/DAG-types) on a fixed set of parties/labs.
Morphisms represent \emph{refinement} or \emph{forgetting} of causal detail.
\end{definition}

Note that for two parties $A,B$, $\mathcal{O}$ may be the two-point poset $\{A\prec B,\;B\prec A\}$.
For many parties, $\mathcal{O}$ can be taken as (a suitable subcategory of) DAG-types/partial orders.

\section{Local intuitionistic semantics inside a context}

The first ingredient is deliberately conservative: \emph{inside any fixed context $c$ we use
ordinary intuitionistic semantics.} This makes ``indeterminate'' a first-class notion even before
any seven-valued layer is introduced.

\paragraph{Why intuitionistic logic:}
Intuitionistic logic is a refinement of classical logic in which truth is understood as
\emph{warranted assertibility} (provability) rather than as an a priori bivalent property.
In particular, one does not assume the law of excluded middle ($\varphi\vee\neg\varphi$) holds
for arbitrary propositions. As a result, it is meaningful---and often natural---that, relative to a
given body of information or a fixed ``context,'' neither $\varphi$ nor $\neg\varphi$ is warranted.
We adopt intuitionistic semantics here because our central notion of \emph{indeterminate} is precisely
of this form: before introducing any cross-context bookkeeping, a single context may already leave
order-claims or causal claims undecided. The seven-valued layer introduced later does not replace
intuitionistic reasoning; it records, across multiple contexts, whether a proposition is supported
somewhere, refuted somewhere, and/or indeterminate somewhere \cite{Kripke1965,TroelstraVanDalen1988,Isham2010ToposMethods}.

\begin{definition}[Local forcing (Kripke-style)]
Fix a refinement preorder $\le$ on contexts (implicit in $\mathcal{O}$).
For propositions $\varphi$ about a process (order, signalling, separability, etc.), write
$c\Vdash \varphi$ for ``$\varphi$ is warranted in context $c$'' and also require monotonicity:
\[
c\Vdash\varphi \ \text{and}\ c\le d \quad\Rightarrow\quad d\Vdash\varphi.
\]
Define intuitionistic negation by
\[
c\Vdash \neg\varphi \quad\Longleftrightarrow\quad (\forall d\ge c)\ (d\not\Vdash \varphi).
\]
\end{definition}

One should note that indeterminacy already exists locally.
In a fixed context $c$, it may happen that neither $c\Vdash \varphi$ nor $c\Vdash \neg\varphi$.
This is the intuitionistic notion of \emph{undecided}. We will fold this together with
``not appropriately posed in $c$'' under the single label \emph{indeterminate} later.

\section{Seven-valued contextual semantics across contexts}

Now we pass from \emph{local} intuitionistic truth to a \emph{contextual summary across a family of
contexts} $\mathcal{C}\subseteq\mathrm{Ob}(\mathcal{O})$ relevant to a given discussion/experiment.

\subsection*{Three meta-predicates: supported, refuted, indeterminate}
\begin{definition}[Supported/refuted/indeterminate across $\mathcal{C}$]
For a proposition $\varphi$ define:
\[
\mathbf{T}(\varphi) \;:\Longleftrightarrow\; \exists c\in\mathcal{C}\ (c\Vdash \varphi),
\qquad
\mathbf{F}(\varphi) \;:\Longleftrightarrow\; \exists c\in\mathcal{C}\ (c\Vdash \neg\varphi),
\]
\[
\mathbf{I}(\varphi) \;:\Longleftrightarrow\; \exists c\in\mathcal{C}\ (\varphi \text{ is indeterminate in } c).
\]
Here ``$\varphi$ is indeterminate in $c$'' is an umbrella that includes:
\begin{itemize}[leftmargin=2em]
\item intuitionistic undecidedness in $c$ (neither $\varphi$ nor $\neg\varphi$ is forced), and
\item cases where $\varphi$ is not appropriately posed/typed in $c$ (recorded simply as \emph{indeterminate}).
\end{itemize}
\end{definition}

\subsection*{The seven values}
Define the contextual value of $\varphi$ as the nonempty subset
\[
v(\varphi)\ :=\ \{\,X\in\{\mathbf{T},\mathbf{F},\mathbf{I}\}\ :\ X(\varphi)\ \text{holds}\,\}.
\]
There are exactly seven nonempty possibilities:
\[
\mathbf{T},\ \mathbf{F},\ \mathbf{TF},\ \mathbf{I},\ \mathbf{TI},\ \mathbf{FI},\ \mathbf{TFI}.
\]

We emphasize that this is intuitionistic elaboration, not replacement.
All reasoning about $\varphi$ and $\neg\varphi$ remains intuitionistic \emph{within} contexts.
	The seven-valued output is a \emph{meta-level classifier across contexts}; it is not (at this stage)
a truth-functional semantics for connectives.

\subsection*{Why this helps for causal order}
For causal-order propositions (e.g.\ $\varphi=$ ``$A\prec B$''), the classifier separates three effects:
\begin{enumerate}[leftmargin=2em]
\item \emph{Context dependence:} $\mathbf{TF}$ means supported in some definite-order contexts and refuted in others.
\item \emph{Order-indeterminacy:} the presence of $\mathbf{I}$ means that at least one relevant context leaves
order claims indeterminate (including ``not appropriately posed/typed'' in that context).
\item \emph{Cross-context conflict plus indeterminacy:} $\mathbf{TFI}$ distinguishes ICO-like discourse from
mere ignorance about which definite order obtained.
\end{enumerate}

\section{Failure-to-glue: causal separability as a global-section property}

The second ingredient is structural: make ``a global definite-order story exists'' into a gluing statement.

\subsection*{A presheaf of definite-order models}
\begin{definition}[Presheaf of admissible models over $\mathcal{O}$]
Let $\mathsf{Adm}:\mathcal{O}^{op}\to \mathsf{Conv}$ be a presheaf into convex sets and affine maps, where:
\begin{itemize}[leftmargin=2em]
\item $\mathsf{Adm}(c)$ is the convex set of admissible process-descriptions compatible with definite context $c$,
\item restriction maps implement forgetting/refining causal constraints along morphisms in $\mathcal{O}$.
\end{itemize}
\end{definition}

\begin{definition}[Global section (definite-order globalisation)]
A \emph{global section} is a compatible choice of elements across all contexts in $\mathcal{C}$
that agrees under restriction on overlaps/refinements (in the standard presheaf sense).
\end{definition}

\subsection{The key structural slogan}
\begin{proposition}[Causal separability as globalisability (guiding principle)]
When $\mathsf{Adm}$ is chosen so that its sections encode definite-order explanations, the statement
\[
\text{``the overall behaviour is a classical mixture of definite orders''}
\]
corresponds to the existence of an appropriate global section (or, equivalently, a factorisation through
a canonical colimit/copower representing mixtures of definite-order contexts).
Failure of such a global section corresponds to a \emph{failure-to-glue} of definite-order descriptions.
\end{proposition}

This recasts the familiar convex-decomposition criterion into a \emph{geometric/semantic} statement:
\emph{no global definite-order section exists.} It also aligns ICO with the ``no global section'' paradigm
used elsewhere for contextuality\cite{AB}.

\section{Toy synthesis: how the two layers work together}

\begin{example}[Two parties: separating dependence, conflict, indeterminacy, and gluing]
Let $\mathcal{C}=\{c_{AB},c_{BA},c_{\mathrm{ico}}\}$ where $c_{AB}$ and $c_{BA}$ are definite-order contexts,
and $c_{\mathrm{ico}}$ is a context intended to speak about an order-indefinite global description
(e.g.\ a higher-order wiring/resource).

Consider $\varphi=$ ``$A\prec B$''.
\begin{enumerate}[leftmargin=2em]
\item Typically $c_{AB}\Vdash\varphi$ and $c_{BA}\Vdash\neg\varphi$, yielding $\mathbf{TF}$.
\item In an ICO-like context, $\varphi$ may be indeterminate in $c_{\mathrm{ico}}$, yielding $\mathbf{TFI}$.
\end{enumerate}
Now consider $\chi=$ ``the process is causally separable'' (i.e.\ explainable as a classical mixture of
definite orders). Then:
\begin{enumerate}[leftmargin=2em]
\item If $\chi$ holds, $\mathsf{Adm}$ has a global section capturing a mixture of definite-order sections.
\item If $\chi$ fails, there is no such global section: definite-order models do not glue.
\end{enumerate}
Thus $\mathbf{I}$ (semantic indeterminacy) and failure-to-glue (structural obstruction) reinforce each other:
the first prevents collapsing ICO talk into ``unknown order,'' and the second prevents collapsing it into
``a hidden mixture of definite orders.''
\end{example}

\section{How this perspective differs from existing ICO approaches}

\subsection*{Making ``no definite order'' a canonical gluing obstruction}
Standard ICO work often characterises definiteness by convex decompositions (causal separability) or by
inequality tests\cite{OCB} or causal-witness tests\cite{AraujoWitness}. The present approach re-expresses the core intuition as a single geometric idea:
\begin{center}
\emph{Definite-order explanations fail because definite-order sections do not glue into a global section.}
\end{center}
This places ICO side-by-side with well-developed ``no global section'' methodologies and invites the use of
standard sheaf/presheaf technology (including obstruction theory).

\subsection*{A disciplined semantic metalanguage that separates three phenomena}
The seven-valued layer records, in a compact and intuitionistically grounded way, whether a proposition is:
supported somewhere, refuted somewhere, and/or indeterminate somewhere.
This yields distinctions that are often conflated in informal ICO discussions:
\begin{itemize}[leftmargin=2em]
\item \emph{Context dependence across definite orders} ($\mathbf{TF}$) vs
\item \emph{genuine order-indeterminacy} (presence of $\mathbf{I}$) vs
\item \emph{simultaneous conflict and indeterminacy} ($\mathbf{TFI}$).
\end{itemize}
Crucially, \emph{indeterminate} here is not ``we do not yet know''; it is a context-sensitive
intuitionistic status (including ``not appropriately posed/typed'') that persists even in a fully specified model.

\subsection*{A bridge between higher-order causal categories and contextual reasoning}
Higher-order causal categories supply the \emph{object language} for admissible composition (an internal logic
of causal wiring). The present seven-valued + gluing synthesis supplies a \emph{meta-semantics} for statements
\emph{about} causal order and its (in)definiteness, indexed by causal contexts/types.
This separation between an internal compositional calculus and an external contextual semantics is conceptually clean and, to our knowledge, not usually presented in this explicit form in the ICO literature. 

\clearpage
\bigskip
\begin{center}
\large\bfseries Part II\\
\medskip
Parametric Time, Stochastic Spin Networks, and Quantum Gravity
\end{center}
\bigskip
\section{Part II: Parametric Time, Stochastic Spin Networks, and Quantum Gravity}

Part~I was deliberately operational: it treated \emph{definite} causal structures as contexts and made
``definite-order explainability'' a \emph{gluing} property (existence of global sections), while the
seven-valued classifier provided a compact contextual semantics that separates \emph{support}, \emph{refutation},
and \emph{indeterminacy} across contexts. In this Part~II we sketch how the same semantic--structural
package can be used as a guiding language for quantum gravity, under the thesis that \emph{time is
fundamentally a parametric ordering variable} rather than a geometric datum.

\subsection{Parametric time as order, not geometry}

A recurring lesson of canonical quantum gravity is that ``time'' should not be treated as a primitive
external coordinate. We adopt a sharper working hypothesis:
\begin{quote}
There exists a global \emph{parametric} time $\tau$ that orders physical updates, while familiar geometric
(or clock) time is emergent and relational.
\end{quote}
This is congenial to stochastic-mechanical foundations of quantum theory: one starts with an underlying
$\tau$-ordered stochastic dynamics and recovers quantum amplitudes and effective ``timeless'' constraints
as equilibrium conditions. On this view, indefinite causal order (ICO) should be understood primarily as
\emph{indefiniteness of the parametric ordering of effective interventions/events}, not as a primitive denial
of all ordering.

Two remarks prevent ambiguity.

\emph{(i) Timeless spin networks versus stochastic spin networks.} Standard spin networks (as used in canonical loop quantum gravity) are kinematical objects for \emph{space}: they represent quantum states of a three-geometry and, by themselves, carry no time parameter. Introducing a \emph{stochastic} spin-network dynamics adds a genuine ordering parameter $\tau$ governing local updates (flips/recouplings/transport of the underlying degrees of freedom). In a continuum limit, one should therefore expect---at least at the first stage---an emergent \emph{geometric space} together with an evolution/ordering in $\tau$, i.e.\ a family of spatial geometries parameterised by $\tau$, rather than a fundamental four-dimensional spacetime with geometric time as part of the primitive structure. Recovering an effectively relativistic spacetime description then becomes a further, genuinely emergent step. \cite{NandiGhosePetruccione2025}

\emph{(ii) Hilbert space and ``spacetime superposition.''} In stochastic quantization and related stochastic-mechanical foundations, Hilbert space is not assumed as a primitive arena: when a Hilbert-space representation appears, it is reconstructed from an underlying probabilistic dynamics (typically in an equilibrium or stationary sector). Consequently, ``superpositions of spacetimes'' in the conventional sense are not taken as fundamental ingredients here. The relevant notion of ICO is instead the possible failure of a definite \emph{parametric} ordering for coarse-grained, relational interventions, even though the microscopic update process is globally ordered by $\tau$. \cite{NandiGhosePetruccione2025}

\subsection{A stochastic spin-network dynamics with helicity data}

Let $\Gamma(\tau)$ be a stochastic process on a configuration space $\mathcal{S}$ of spin networks enriched
by helicity labels. Concretely, a configuration may be written schematically as
\[\Gamma \equiv (G,\{j_e\},\{\iota_v\},\{h_e\}),\]
where $G$ is a graph, $j_e$ are representation labels on edges, $\iota_v$ are intertwiners at vertices, and
$h_e\in\{+,-\}$ (or an appropriate finite set) encodes a chiral/helicity datum.
Assume $\tau$-evolution is given by a local-move dynamics (edge flips, recouplings, splittings/mergings)
with a Markov generator $\mathcal{L}$ acting on probability densities $\rho(\Gamma,\tau)$:
\begin{equation}
\partial_\tau \rho(\Gamma,\tau)=\mathcal{L}\,\rho(\Gamma,\tau).
\label{eq:FP}
\end{equation}
In jump-process form,
\begin{equation}
(\mathcal{L}f)(\Gamma)=\sum_{\Gamma'\sim\Gamma} r(\Gamma\to\Gamma')\,[f(\Gamma')-f(\Gamma)],
\label{eq:master}
\end{equation}
with rates $r(\Gamma\to\Gamma')$ depending on local labels.

In this setting a Wheeler--DeWitt-type ``timeless'' constraint may be interpreted as an equilibrium
condition with respect to $\tau$-evolution: physical states correspond to stationary solutions
\begin{equation}
\mathcal{L}\,\rho_{\mathrm{phys}}=0,
\label{eq:stationary}
\end{equation}
together with an amplitude-level construction (Nelson-type or related) that builds a complex functional
$\Psi$ from $\rho$ plus a phase-like variable. The detailed identification with a specific operator
constraint $\widehat{H}\Psi=0$ is model-dependent; for the causal-order discussion it suffices that
``physicality'' is tied to stationarity/equilibrium in $\tau$ \cite{NandiGhosePetruccione2025}.

\paragraph{Parametric versus geometric time}
In the stochastic-spin-network approach, $\tau$ is an intrinsic ordering parameter for microscopic updates.
Interpreting the Wheeler--DeWitt condition as an equilibrium/stationarity requirement means that the
\emph{state} (or distribution) becomes $\tau$-independent (schematically, $\mathcal{L}\rho_\ast=0$), not that
the underlying $\tau$-ordered stochastic process ceases to evolve. Thus ``geometric time is frozen'' refers
to the absence of a fundamental geometric time variable in the equilibrium description, while $\tau$ persists
as an auxiliary ordering parameter whose detailed evolution is typically unobservable in equilibrium expectation
values. In this setting, indefinite causal order concerns the $\tau$-ordering of \emph{coarse-grained relational}
interventions/events, which may remain indeterminate even when the underlying dynamics is $\tau$-ordered and
stationary in distribution.

\subsection{Effective events and interventions are generically $\tau$-delocalised}

In a quantum-gravity setting, an ``event'' is not a primitive coordinate point. Rather, it is identified
\emph{relationally} as the emergence of a stable record: a clock reading, a detector click, a macroscopic
pointer value, or a robust correlation between subsystems extracted by coarse-graining. In a spin-network
language this means that an event is specified by a coarse-grained predicate on configurations---for example,
``a chosen clock degree of freedom takes a value in a window $\Delta$ while a detector degree of freedom in a
given subgraph registers a transition.'' Here the ``clock'' is an internal relational variable $T(\Gamma)$ used to time-stamp events; it is distinct from the parametric ordering variable $\tau$ and also from any emergent geometric time that
appears only in semiclassical regimes.

One convenient way to make this explicit is to treat each event $A$ as a condition $E_A(\Gamma)=1$ on the
microscopic configuration $\Gamma$, possibly involving an internal clock functional $T(\Gamma)$.
The \emph{parametric time of occurrence} of the corresponding effective record is then a (generally random)
hitting time,
\begin{equation}
\tau_A \;:=\; \inf\{\tau:\ E_A(\Gamma(\tau))=1\},
\end{equation}
and similarly for $B$. Even when the underlying update process is globally ordered by $\tau$, the induced
distributions of $\tau_A$ and $\tau_B$ need not be sharp, and in deep quantum-gravity regimes they are
expected to be broad and overlapping. In such situations both order-relations can occur with non-negligible
weight, e.g.\ $\mathbb{P}(\tau_A<\tau_B)>0$ and $\mathbb{P}(\tau_B<\tau_A)>0$, and the effective order of
events is not well captured by a single definite ``$A$ before $B$'' narrative.

A complementary (and often more convenient) phenomenological description is to model an effective
\emph{intervention} $A$ as a $\tau$-smeared coupling,
\begin{equation}
A\;\equiv\;\int d\tau\, f_A(\tau)\,\mathcal{A}_\tau,
\label{eq:smearedA}
\end{equation}
where $\mathcal{A}_\tau$ denotes the elementary interaction at parametric time $\tau$ and $f_A$ is an
envelope that encodes the $\tau$-delocalisation of the record/intervention (narrow in semiclassical regimes;
broad in deep quantum-gravity regimes). Likewise,
\begin{equation}
B\;\equiv\;\int d\tau\, f_B(\tau)\,\mathcal{B}_\tau.
\label{eq:smearedB}
\end{equation}
In a relational setting, the envelopes $f_A,f_B$ can be understood as induced by the clock functional and
the coarse-graining that defines the events: they summarise ``when, in $\tau$,'' the corresponding coupling
effectively acts.

When $f_A$ and $f_B$ overlap (or when the $\tau$-location of the event, defined by relational clock/record conditions, is correlated with different branches of the coarse-grained history), there may be no sharp
fact of the matter about whether $A$ occurs before $B$ in the \emph{effective} parametric order, even though
$\tau$ itself orders microscopic updates. The point is not that $\tau$ fails to order the underlying dynamics;
rather, the \emph{order-proposition about the coarse-grained events} can remain indeterminate because the events
are defined only relationally and are delocalised in $\tau$.
\subsection{ICO as indeterminacy of parametric order}

To import the Part~I framework, we take \emph{contexts} to be definite causal-order types in parametric time.
For two effective interventions, two basic contexts are
\[c_{AB}:\ A\prec_\tau B,\qquad c_{BA}:\ B\prec_\tau A,\]
and for many parties one uses DAG-types encoding one-way signalling constraints in $\tau$.
Let $\mathcal{O}_\tau$ be the resulting category/poset of definite $\tau$-orders.

Over $\mathcal{O}_\tau$ we consider a convex presheaf $\mathsf{Adm}_\tau: \mathcal{O}_\tau^{op}\to\mathsf{Conv}$
of admissible definite-$\tau$ explanations of the induced operational statistics. The Part~I slogan becomes:
\begin{quote}
\emph{Definite-order explainability in $\tau$ corresponds to the existence of a global section of
$\mathsf{Adm}_\tau$; ICO in the parametric sense is a failure-to-glue of definite-$\tau$ sections.}
\end{quote}

The seven-valued contextual classifier then supplies the minimal semantic bookkeeping for order-propositions.
For instance, for the proposition $\varphi=$ ``$A\prec_\tau B$'', one can distinguish:
(i) variation across definite-$\tau$ contexts (support in $c_{AB}$ and refutation in $c_{BA}$), from
(ii) genuine \emph{indeterminacy} of $\varphi$ in an order-indefinite effective description induced by
$\tau$-delocalisation, and from (iii) their coexistence. Crucially, the indeterminate status here is not
``mere ignorance'': it persists even when the underlying $\tau$-ordered microscopic dynamics is fully
specified, because it arises from coarse-grained relational localisation.

\subsection{Relation to existing ICO strands and a concrete programme}

This parametric-time reading sharpens several debates highlighted in the ICO literature. In particular, it separates proposals tied to geometric-time superpositions of spacetime causal structure from the present parametric-time stance, where ordering is carried by $\tau$ and geometry is emergent. Switch-like
interference can be viewed as coherent control of effective $\tau$-ordering; process-matrix descriptions can
be read as operational summaries that need not commit to a global definite-$\tau$ section; and superpositions
of causal structure in quantum gravity can be interpreted as branch-dependent emergent geometry built over
an underlying $\tau$-ordered dynamics (cf.\ the gravitational-switch perspective in \cite{Zych2019}).

A minimal research programme is then:
\begin{enumerate}[leftmargin=2em]
\item Construct a finite toy dynamics (few spin-network configurations and local moves) and define two
relational interventions $A,B$ with envelopes $f_A,f_B$. Compute the induced operational statistics and test
whether $\mathsf{Adm}_\tau$ admits a global section.
\item Formulate a quantitative ``degree of non-gluability'' as distance to the gluable (global-section) subset
in the relevant convex structure, and study its behaviour under composition and coarse-graining.
\item Analyse how helicity/chirality labels affect interference between effective orderings and whether they
yield robust signatures of parametric ICO.
\end{enumerate}

Part~II is therefore not offered as a finished model of quantum gravity, but as a precise proposal for how
the Part~I semantic--structural framework can be used to formulate and compare notions of ICO when time is
treated as an ordering parameter rather than a background geometry.

\section{Concluding Remarks}

Part~I recast causal separability as existence of a global section of a convex presheaf over the poset/category of definite causal orders, yielding a canonical ``failure-to-glue'' criterion that is compositional by construction \cite{AB,Wilson,KU}. The seven-valued (intuitionistic) contextual summary then provided a compact semantics for causal-order propositions, distinguishing (i) variability across definite-order contexts, (ii) genuine indeterminacy in order-indefinite contexts, and (iii) their coexistence.

Part~II proposed a quantum-gravity reading of the same framework in which the fundamental notion of time is
a parametric ordering variable $\tau$, while ``events'' are relational, coarse-grained interventions that may be
delocalised in $\tau$. This perspective is motivated by a stochastic-quantization (stochastic-mechanical) approach
to spin-network dynamics, rather than canonical quantization: Hilbert space and superpositions of spacetime are not
taken as primitive, and the Wheeler--DeWitt condition is interpreted as an equilibrium/stationarity constraint on the
underlying $\tau$-ordered update process. In this setting, indefinite causal order is naturally interpreted as
indeterminacy of the \emph{parametric} order of effective interventions, even when the microscopic dynamics is globally
ordered by $\tau$; and ``no hidden definite order'' becomes a precise non-gluing obstruction for definite-$\tau$
explanations. This also helps separate switch-like order interference from genuinely gravitational scenarios
(cf.\ the ``gravitational switch'' viewpoint in \cite{Zych2019}) by asking the same structural question:
do definite-order contexts glue?

Taken together, the two parts provide a unified semantic--structural language for navigating the ICO
landscape without conflating distinct notions of ``indefinite order.'' On the one hand, decomposition-based
definitions ask whether the observed behaviour admits a definite-order explanation, possibly as a convex
mixture of definite orders---equivalently, whether definite-order descriptions can be glued into a global
section over the chosen context base. On the other hand, witness-based methods provide practical tests,
under stated modelling assumptions, for certifying the failure of such definite-order explanations by
separating causally nonseparable processes from the causally separable set \cite{AraujoWitness}. The present
framework makes the common structure behind these approaches explicit while keeping their assumptions and
scopes distinct.

\section{Acknowledgements}
I thank ChatGPT (OpenAI) for assistance with language polishing and with the organization of ideas and notation during the preparation of this manuscript. However, I am solely responsible for the contents.

\end{document}